\documentclass[aps,prd,preprint,tightenlines,superscriptaddress]{revtex4}
\usepackage{epsfig}
\usepackage{graphicx}
\usepackage{psfrag}
\usepackage{amsmath,amssymb}
\usepackage{colordvi}
\usepackage{color}
\usepackage{amsfonts}
\usepackage{enumerate}
\usepackage{slashed,bm}

\begin{document}

\title{Parton Physics from Large-Momentum Effective Field Theory}
\author{Xiangdong Ji}
\affiliation{INPAC, Department of Physics and Astronomy, Shanghai Jiao Tong University, Shanghai, 200240, P. R. China}
\affiliation{Maryland Center for Fundamental Physics, Department of Physics,  University of Maryland, College Park, Maryland 20742, USA}
\date{\today}
\vspace{0.5in}
\begin{abstract}

Parton physics, when formulated as
light-front correlations, are difficult to study non-perturbatively, despite the promise
of light-front quantization. Recently an alternative approach to partons have been
proposed by re-visiting original Feynman picture of a hadron moving at asymptotically large
momentum. Here I formulate the approach in the language of an effective field theory for a large hadron momentum
$P$ in lattice QCD, LaMET for short. I show that using this new effective theory, parton properties, including 
light-front parton wave functions, can be extracted from lattice observables in a
systematic expansion of $1/P$, much like that the parton distributions can be extracted from the
hard scattering data at momentum scales of a few GeV.

\end{abstract}

\maketitle

One of the greatest simplifications in describing physics of hadron scattering
at high energy, such as production of Higgs bosons at the Large Hadron Collider (LHC)~\cite{science}, 
is the parton model introduced by R. Feynman~\cite{Feynman:1969ej}. According to this, a fast moving hadron, such as proton, can be viewed
as a beam of noninteracting quarks and gluons (partons) characterized by their momentum density,
$q(x)$ and $g(x)$, where $x$ is the fraction of the longitudinal momentum carried by the parton, $x= k^z/P^z$ with
the hadron momentum $P^z \rightarrow \infty$. Then the hard-scattering cross sections involving the hadrons can be
calculated as the convolution of the basic parton scattering cross sections $\hat \sigma$ and parton densities.
In the fundamental theory of strong interactions, quantum chromodynamics (QCD), this simple picture can
be justified by the so-called factorization theorems~\cite{Mueller:1989hs,Sterman:1994ce,Collins:2011zzd}. The only corrections
introduced by quantum field theory
is that the parton densities are scheme and scale dependent, and the latter can be studied through renormalization 
group equations~\cite{Altarelli:1977zs}. The scheme and scale dependence
of the parton densities are of course cancelled by the similar dependences of parton scattering cross sections,
leaving the physical quantities invariant under the perturbative definitions
of partons.

While parton scattering cross sections can be computed in QCD perturbation theory
thanks to asymptotic freedom, the parton densities are intrinsically non-perturbative. 
As mentioned above, Feynman's definition of parton density was made in the infinite momentum frame (IMF), in which 
the parent hadrons have an infinite momentum~\cite{Feynman:1969ej}. This seems to be a mathematical limit difficult
for intuitive understandings. The exact notion of the infinite-momentum limit can actually be understood by
boosting Feynman diagrams in perturbation theory, as was done by Weinberg~\cite{Weinberg:1966jm}. Over the years,
however, one found that it is convenient to formulate the parton
density in the formalism of light-cone correlation function~\cite{Drell:1969ca}: Introduce the light-cone
coordinates,
\begin{equation}
      \xi^\pm = \frac{1}{2}(\xi^0\pm \xi^3) \ ,
\end{equation}
and similarly for other four-vectors, where $\xi^\mu$ $(\mu=0,1,2,3)$ is the space-time coordinates 
(reserving $x$ for the momentum fraction) and the hadron
is moving in the 3 or $z$-direction. The parton density is now calculated as the matrix elements
of the non-local correlator~\cite{Collins:1981uw},
\begin{eqnarray}
   q(x, \mu^2) &=& \int \frac{d\xi^-}{4\pi} e^{-ix\xi^-P^+}  \langle P|\overline{\psi}(\xi^-)
   \gamma^+ \\
   && \times \exp\left(-ig\int^{\xi^-}_0 d\eta^- A^+(\eta^-) \right)\psi(0) |P\rangle \nonumber \ ,
\end{eqnarray}
where $\psi$ the quark Dirac field, $A^\mu$ is the gluon potential,
$g$ is the strong coupling, $\mu^2$ is the renormalization scale.
Here the infinite boost factors all disappeared, and IMF physics is reflected entirely through
the boosted quark and gluon operator. The above matrix element is now independent of the hadron
external momentum $P^\mu$! The price one pays,
however, is that the parton density no longer represents a equal-time correlation, as 
was originally defined in the IMF by Feynman.

Although a significant progress is made by getting rid of the infinite boost, the
light-cone correlation functions by no means are easy to calculate. For instance, to use Wilson's
lattice QCD method~\cite{Wilson:1974sk}, which is intrinsically Euclidean, one has to get rid of the real
time dependence. One has to
Taylor-expand the separation between the quark fields and obtain the local operators with multiples
of derivatives, whose matrix elements are related to the moments of parton distributions~\cite{negele}. However, measuring
operators matrix elements with many derivatives is intrinsically noisy, and as such, one can only evaluate
the first few moments. Over the years, people have invented approaches to overcome
this difficulty~\cite{Aglietti:1998mz,Liu:1999ak,Detmold:2005gg}, but
none has been promising enough for realistic numerical simulations.

An alternative approach that has been advocated strongly by some is light-front 
quantization~\cite{Ji:1992ku,Burkardt:1995ct,Brodsky:1997de},
related to the form of the time-independent perturbation theory in
the IMF~\cite{Weinberg:1966jm}. Actually the light-front formulation of a dynamical theory goes back
more than half a century to Dirac's original paper~\cite{Dirac:1949cp}.
The formalism goes roughly as follows: Consider $\xi^+$ as the new time and introduce the equal "time"
commutators to quantize the field theory. The fields at $\xi^+=0$ have canonical plane-wave expansions in terms of
the Fock particles. The new hamiltonian is $\hat H_{\rm LC}= \hat P^-$, which
can be used to develop time-independent perturbation theory as in non-relativistic quantum mechanics.  One can then go on
to determine the spectrum of $P^-$ and eigenstates by solving the non-perturbative Schr\"odinger-like equation,
\begin{equation}
      \hat P^-|P\rangle = (M^2/2P^+)|P\rangle \ ,
\end{equation}
where eigenvalue $M^2$ is the hadron mass, and $P^+$ is the light-front momentum.
In QCD, it is most convenient to work with $A^+=0$ gauge, the Fock particles are partons.
The hadron states can be expressed in terms of the Fock expansion~\cite{Brodsky:1997de},
\begin{equation}
             |P\rangle = \sum_{n\alpha, \lambda_i} \int \Pi_i \frac{dx_id^2k_{\perp i}}{\sqrt{2x_i}(2\pi)^3} \psi_{n\alpha}(x_i, k_{\perp i}, \lambda_i) \left|n\alpha
             : x_iP^+, k_{\perp,i},\lambda_i \right\rangle \ ,
\end{equation}
where $n$ is the number of partons, $\lambda_i$ are helicity labels. All partons have the longitudinal fraction $x_i$,
and transverse momentum $k_{\perp, i}$. Index $\alpha$ sums over possible amplitudes $\psi_{n\alpha}$ for a given parton number and
parton helicities. The examples of the proton and meson Fock states can be found in Ref. \cite{Ji:2002xn,Ji:2003yj}.
Once we have the light-front wave functions, one can calculate any parton physics observables of interest.

Solving a field theory on the light-cone front is notoriously difficult, despite the many nice features of the theory
such as the vacuum becomes "trivial", and one more generator of the Lorentz group
becoming kinematical, etc.~\cite{Brodsky:1997de}. Apart from two-dimensional theories and
perturbative expansion in terms of the number of Fock particles, no systematic
approximation has been found for non-perturbative calculations in 3+1 dimension. Unlike the ordinary
formulation of asymptotic field theories whose static properties can be simulated on Euclidean lattices,
no such formalism exists for light-front theories. The difficulty might be related to the fact that they 
are intrinsically Minkowskian. A hybrid formulation of
the light-front theories in terms of the transverse lattice has been proposed and explored~\cite{Bardeen:1979xx},
but so far it has not lead to successful simulations.

In recent publications, a new approach to calculating parton distributions using Euclidean
lattice QCD has been proposed~\cite{Ji:2013fga, Ji:2013dva, Xiong:2013bka,Hatta:2013gta,Lin:2014zya}.
This approach essentially goes back to the original definition
of the parton densities by Feynman, i.e., starting with the ordinary momentum distribution $n(\vec{k}, P^z)$ 
related to the space correlation function of the quark fields in a hadron. This quantity
is calculable using ordinary lattice QCD. However,  it depends
on the hadron momentum $P^z$. The Feynman distribution is obtained in the limit of
$P^z\rightarrow \infty$ limit, i. e,
\begin{equation}
                q(x) \sim \lim_{P_z\rightarrow\infty} \int d^2k_\perp dk^z n(\vec{k},P^z)\delta (x-k^z/P^z) \ .
\end{equation}
The lattice simulations, however, cannot provide an infinite $P^z$ result directly.  
How can one then recover the $P^z=\infty$ limit from a moderately-large-$P^z$ result that might
be possible on the lattice?  An added subtlety is that field theory usually has ultraviolet (UV) divergences.
In a typical Feynman diagram, the order of regularization UV divergences vs. $P^z\rightarrow \infty$ does not
commute with each other. While defining the parton distribution requires $P^z$ to be much larger than the cut-off scale,
the lattice simulations only makes sense when the UV cut-off is much larger than the hadron momentum.

The solution to the above problems have been discussed in Refs. \cite{Ji:2013dva}: $n(\vec{k}, P^z)$
calculable in lattice QCD can be matched to the parton density $q(x)$ through perturbatively calculable
multiplicative factor $Z$ plus higher-order corrections in powers of $(1/P^z)^2$.
The approach can be generalized to any other light-front quantities and can be formulated in terms of
an effective field theory of lattice QCD in the presence of one (or more) large momentum scale $P^z$.
The situation is very similar to that of the heavy quark effective field theory (HQET), which by now
has been well-understood and widely used~\cite{Manohar:2000dt}.  Here I discuss in some detail
this new {\it large momentum effective field theory} (LaMET),  which allows parton physics to be calculated
in lattice QCD at finite $P^z$ in a systematic approximation.

Let us first review how HQET works. Consider a physical system with one heavy quark, such as
$B^-$ meson with the $b\bar u$ quark content. A general physical observable $O$ of the system,
such as the weak decay constant, will depends on the $b$-quark mass parameter $m_b$.
Because of the asymptotic freedom, $O(m_b)$ for a large $m_b>\!>\Lambda_{\rm QCD}$, where $\Lambda_{\rm QCD}$
is the QCD scale, can be expanded in terms of powers of $1/m_b$,
\begin{equation}
             O(m_b/\Lambda) = Z(m_b/\Lambda, \Lambda/\mu)o(\mu) + {\cal O}(1/m_b)+ ...
\end{equation}
where we have assumed $O$ depends on some UV cutoff $\Lambda$, which can also be replaced
by a renormalization scale after proper renormaliation in the full theory.
The leading term in the expansion contains the logarithmic dependence of the heavy
quark mass in the matching coefficient $Z$, which has a perturbative
expansion in $\alpha_s$. $o(\mu)$ is a quantity defined in the effective theory,
namely HQET, in which the $b$-quark is effectively infinitely heavy. 
$\mu$ is the renormalization scale in HQET. The sub-leading terms are suppressed
by at least one power of $1/m_b$.

From perturbation theory, it is easy to understand what HQET does. Consider a Feynman diagram
with quark-mass dependence in the external wave functions and internal propagators. In the full theory,
these diagrams are well defined after proper UV regularization. In the effective theory, one simply
takes $m_b\rightarrow \infty$ inside the integrals. This limit shall not change the infrared
property of the quantity under calculation. Thus both $O$ and $o$ in the full and effective theories, respectively,
contains the same non-perturbative physics. However, they do have different UV physics.
And the difference is reflected in the multiplicative matching factor $Z$, which is infrared free
and hence calculable in perturbation theory. The non-trivial aspect here is $Z$
is multiplicative. This can be proved order by order in perturbation theory, just like
proving perturbative renormalizability~\cite{Collins:1984xc}. Of course, we have ignored the complication
that there might be mixings of several operators of the same quantum numbers, and 
there is also a possibility of a convolution between $Z$ and $o(\mu)$ in a certain parameter space.

The important physical observation in applications of HQET is that once $m_b$ is in the perturbation region,
the difference between $O(m_b, \Lambda)$ and its effective theory couner-part $o(\mu)$ is under control, namely we know
how to calculate the matching between them. In certain cases, we need to take into account
of the power corrections which can be categorized and studied in details. Thus to a certain extent,
the physics of a heavy quark is similar to that of an infinitely-heavy one. When can we consider a quark
to be heavy? For some physical observables, the charm quark of mass 1.5 GeV is already a good candidate.
For the bottom quark of mass 4.2 GeV, HQET is expected to work fairly well.

Now we can consider the formulation of LaMET. Consider a lattice (Euclidean) observable $F$ which depends
on a large hadron momentum $P^z$. Using asymptotic freedom, we can systematically
expand the $P_z$ dependence,
\begin{equation}
             F(P^z/\Lambda) = Z(P^z/\Lambda, \Lambda/\mu)f(\mu) + {\cal O}(1/(P^z)^2)+ ... \ . 
\label{Eq:Expansion}
\end{equation}
The quantity $f(\mu)$ is defined in a theory with $P^z\rightarrow \infty$, exactly
as in Feynman's parton model. In other words, $f(\mu)$ is a light-front correlation or parton observable, 
containing all the infrared collinear singularities.
Therefore, Feynman's parton model is in fact an effective theory for the nucleon moving at large momentum!
The important message of the expansion is that it may converge at moderately large $P^z$, allowing
access to quantities defined at infinite $P^z$. The extraction of the parton physics 
can be made more precise by accurately calculating the matching factor $Z$ and higher-order corrections.

The examples of the above expansion have been considered in Ref.\cite{Ji:2013dva, Ji:2013fga}. 
Since the large momentum scale comes from the external states, we do not attempt to define
the effective theory through the lagrangian formalism, just like in the case of perturbative QCD
discussion of factorization theorems.

Momentum dependence of the lattice quasi-observables can be studied through the
renormalization group. Define the anomalous dimension through
\begin{equation}
            \gamma(\alpha_s) = \frac{1}{Z}\frac{\partial Z}{\partial \ln P^z} \ .
\end{equation}
Then one has,
\begin{equation}
            \frac{\partial F(P^z)}{\partial \ln P^z} = \gamma(\alpha_s) F(P^z) + {\cal O}(1/(P^z)^2) \ ,
\end{equation}
up to power corrections. One can sum large logarithms involving $P^z$ using the above equation.

The reason for the existence of the above expansion or effective description is similar to the
existence of HQET. When taking $P^z\rightarrow \infty$ first in $F(P^z)$ before a UV regularization is imposed,
one recovers from $\hat F$ the light-cone operator $\hat f$, by construction. On the other hand, the lattice matrix
element is calculated at large $P^z$, with UV regularization (lattice cut-off) imposed first. Thus the difference
between the matrix elements $f$ and $F$ is the matter of the orders of the limits. This is the standard set-up
for an effective field theory. The different limits do not change the infrared physics. In fact the
factorization in terms of Feynman diagrams can be proved order by order as in the renormalization program.

The above effective theory expansion gives a recipe to study parton physics in lattice QCD: 
First, start from a particular parton observable $\hat f$ which is an operator made of 
light-cone fields. Then construct an
Euclidean version $\hat F$ which, under an infinite Lorentz boost, goes to $\hat f$. Calculate the lattice matrix element
of $\hat F$ in a hadron with large momentum $P^z$ and use Eq. \ref{Eq:Expansion} to extract the parton physics $\hat f$.
Of course, matrix element $F$ depends on $P^z$ as well as all the lattice UV artifacts. All these physics will be
captured in the matching factor $Z$.

While the previous works have shown examples how LaMET work for parton densities, here I would like to
consider the light-front wave functions itself. One can calculate any parton physics once the complete 
light-front wave functions are known, and this is the main object the light-front quantization is after. 
For simplicity, let us consider the $\pi^+$ meson, although a similar discussion applies for a proton.
The leading light-front wave function is related to the matrix element,
\begin{equation}
  \langle 0|\bar d(0)\gamma^+\gamma_5 u(\xi^-,\xi_\perp)|\pi^+(P)\rangle \ , 
\label{Eq:meson}
\end{equation}
where $u$ and $d$ are up and down quark fields in coordinate space, defined in 
which is defined in the $A^+=0$ gauge. In fact all the Fock wave functions in the light-front theories
can be expressed in terms of light-front correlations between the hadron state and QCD vacuum. 
The gauge-invariant version of the above operator
needs some gauge links along the $\xi^-$ direction to either $\xi^- =+\infty$ or $-\infty$.
A choice of the $\xi^-$ directions shall be made once and for all, and two choices are
physically inequivalent, are related to each other by the complex conjugation~\cite{Ji:2002xn,Ji:2003yj}. 
To ensure complete gauge-invariance, the gauge links at infinity must be connected
by transverse gauge links as the transverse gauge potential does not vanish at $\xi^-=\pm \infty$ in this
singular gauge. One needs to specify how the connections are made to make
the above matrix element uniquely defined. This can be done with a straight-line gauge link 
along $\vec{\xi}_\perp$~\cite{Ji:2002aa}. In the case of many field correlations, 
these end-point links can be chosen in an infinite number of possibilities. 
In a certain limiting process of the hadron momentum 
going to infinity, these end-point links shall not contribute as they are outside of the
physical correlation length, as will be discussed below. 

In LaMET, we need to find a quasi-operator, which under the infinite Lorentz boost will recover
the operator in Eq. \ref{Eq:meson}. We choose, 
\begin{equation}
      \hat F_\pm(z, \xi_\perp) = \bar d (0)W^\dagger(\pm\infty,0;0) \gamma^z\gamma_5 W(\pm\infty,z;\xi_\perp)u(z,\xi_\perp)
\end{equation}
where the all fields are at $\xi^0\equiv t = 0$ and distributed along the $\xi^3\equiv z$ direction, $\gamma^+$ is replaced by
$\gamma^z$. To have gauge invariance, we insert a gauge link for every field, which
goes off to $z=\pm \infty$ along the $z$-direction,
\begin{equation}
        W(\pm\infty, z; \xi_\perp) = P\exp\left(\mp ig\int^{\pm\infty}_z dz'A^z(z', \xi_\perp)\right) \ , 
\end{equation}
where $P$ indicates path ordering. 
In gauges like the covariant one, $\hat F$ is already gauge-invariant. On the lattice, however,
one has to insert a gauge link at large $\pm z$ to connect $W^\dagger(\infty,0;0)$
and $W(\infty,z;\vec{\xi}_\perp)$. We likewise define the end-gauge links as needed in the case
of the light-front quantization. Again, the result shall be independent of these end-point links 
as the physical correlation length is finite. 

At this point, it is instructive to compare the extraction of parton physics from high-energy scattering
with that of LaMET. In the high-energy scattering, one starts from scattering cross sections depending on
some large momenta $Q$'s. One performs scale separation and proves factorization theorems, relating the
these cross sections to parton properties; the latter can then be extracted finally from the
experimental data~\cite{Gao:2013xoa}. On the other hand, on a Euclidean lattice, one calculates quasi
observables which depend on the large hadron momentum $P^z$. One can use LaMET to perform scale separation,
and relating these quasi obervables to the parton physics. There is a complete analogy here.
Thus, simply put, LaMET allows using lattice "data" to extract parton physics.

Just like the same parton distribution can be extracted from different hard scattering processes, the same
light-front physics can be extracted from different lattice operators. All operators that yield the same light-front
physics form a universality class. The existence of the universality class allows one
exploring different operators $\hat F$ so that a result at finite $P^z$ can be made as close to that at large
$P^z$ as possible. An example of exploring different operators to calculate the total gluon polarization
in the nucleon has been presented in a recent publication~\cite{Hatta:2013gta}.

Finally, let us comment on the relationship between the present approach and light-front quantization. 
In Feynman's original language, the parton physics is described by the matrix elements of type
\begin{equation}
   \langle P_\infty|\hat F|P_\infty\rangle \ ,
\end{equation}
where $\hat F$ is a non-local, time-independent operator and $P_\infty$ indicates a momentum approaching infinity, with possibly a residual momentum
as used in the paper by Weinberg~\cite{Weinberg:1966jm}. On the other hand,
one can write $|P_\infty \rangle
= U_\infty|p_{\rm res} \rangle $, where $p_{\rm res}$ is the residual momentum and $U_\infty$ is a unitary transformation in
the Hilbert space for the infinite boost. Thus, one can write the above quantity
as
\begin{equation}
   \langle p_{\rm res}|U^\dagger \hat FU|p_{\rm res}\rangle \ ,
\end{equation}
with the boost operator acting on $F$. This infinite boost will give rise to a light-front
operator $\hat f\equiv U^\dagger \hat FU$,
\begin{equation}
      \langle P_\infty|\hat F|P_\infty\rangle  = \langle p_{\rm res}|\hat f|p_{\rm res}\rangle \ ,
\end{equation}
where the latter matrix element is independent of the residual momentum. While the present paper directly deals with
the matrix element $\langle P_\infty|\hat F|P_\infty\rangle$ by taking the large $P^z$ limit, the
light-front quantization attempts to calculate the right-hand side of the equation with
light-front wave functions.

The above discussion sheds important light on the space-time picture of the hadron states.
Although one has often draw a round circle of radius about 1fm to represent a coordinate space picture of
a hadron at rest, this runs into problem in quantum mechanics because in a plane wave
state, the hadron can be anywhere in space. One may construct a wave packet to localize a hadron, then
there is no clean separation between the center-of-mass motion and the internal motion in relativistic
theory, unless the hadron is very heavy. More problems appear when considering the spatial picture of
hadron moving with a certain velocity, when one often draws a Lorentz contracted circle. Physical
meaning of the length contraction in a hadron state is often unclear.

\begin{figure*}[!ht]
\begin{center}
\includegraphics[width=0.50\textwidth]{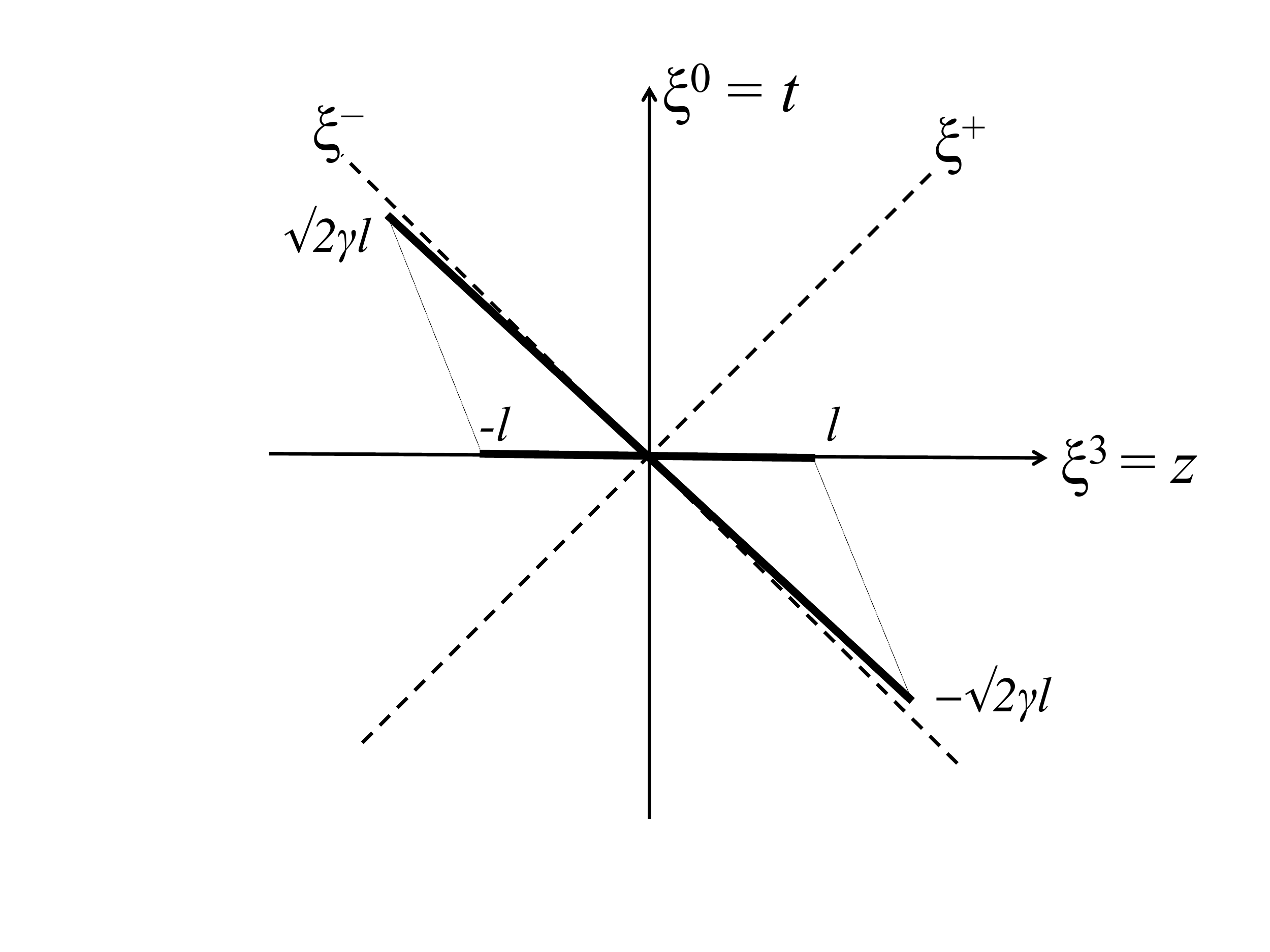}
\end{center}
\caption{The line segment in the space direction $z$ in a frame with a high-momentum hadron is transformed into a line segment along
the light-cone direction in light-front coordinates of the hadron at rest, the length is increased by the boost factor $\gamma$, which goes to infinity 
in the IMF.}
\label{fig:cryosys}
\end{figure*}

One may consider instead the field correlation functions in a plan wave
hadron state. Generally speaking, when the hadron is at rest (zero center-of-mass momentum),
the field correlation functions shall have correlation length about 1fm $\sim \Lambda_{\rm QCD}$
along all the spatial directions. Beyond that, the correlations shall gradually drop to zero.
Now consider a hadron moving with velocity $v$ in the $z$-direction. Introduce the Lorentz factor $\gamma=
1/\sqrt{1-v^2}$. The valence partons will have a fixed fraction of the hadron 
longitudinal momentum, $x=k^z/P^z$. Therefore, the typical longitudinal momentum of the valence 
parton will increase as $xP^z$. Correspondingly, the longitudinal correlation length will go as
$1/xP^z$, inversely proportional to the $\gamma$ factor. This may be considered as the Lorentz 
contraction effect. Thus to calculate the valence parton distributions using LaMET, one has to 
have the longitudinal spatial resolution increases with the nucleon momentum. If one only cares 
about the valence partons, the longitudinal lattice box size can also shrink by a factor of 
$\gamma$. Thus the total number of lattice sites can be similar to that in the transverse directions.

On the other hand, if one cares about the small $x$ partons, one has to have the hadron momentum 
large enough. The smallest $x$ one can reach is $x\sim \Lambda_{\rm QCD}/P^z$. The correlation length
in this case is about $ 1/\Lambda_{\rm QCD}$ which is the normal size of a hadron. For example, at $
x\sim 10^{-4}$, the hadron momentum must be around 3 TeV, and the normal valence partons will have 
correlation length of $10^{-3}$fm. Clearly calculating parton density at such a small $x$ will need
thousands of lattice sites in the $z$-direction.

While one sees a contracted hadron in LaMET, the correlation lengths in the light-front calculations
grow dramatically without bound. This can be seen in Fig. 1, where we have shown a correlation length
$\ell$ along the $z$-axis. In the light-front calculations, the same correlation is now boosted
to along $\xi^-$ direction, increased by a factor of $\gamma$. Thus for valence partons, the correlation length
in the light-front coordinates will be on the order of $\sim 1/\Lambda_{\rm QCD}$. 
For small $x$-partons, the correlation length be as long as $P^z/(M\Lambda_{QCD})$, where $M$ is the mass of
the hadron. Therefore the light-front quantization calculations cannot be limited to a compact space-time 
region, which makes the Monte Carlo simulations difficult.

In summary, we have discussed in this paper the concept of a new effective field theory
for hadrons with large momentum in lattice QCD simulations. The effective theory allows one
to extract the light-front parton physics from practical calculations with hadron momentum 
of order a few GeV. The related expansion involves perturbatively-calculable matching coefficients
and higher-power corrections. This makes the lattice data as useful as the real experimental 
data in studying the bound state properties in QCD.

I thank J. W. Chen, J. Collins, Y. Hatta, H. W. Lin, K. F. Liu, J. W. Qiu, A. Schaefer, F. Yuan, and Y. Zhao for useful
discussions and/or collaborations on the subject. This work was partially supported by the U.
S. Department of Energy via grants DE-FG02-93ER-40762, and a grant (No. 11DZ2260700) from the Office of
Science and Technology in Shanghai Municipal Government, and by National Science Foundation of China (No. 11175114).

\end{document}